\documentclass[runningheads]{llncs}
\usepackage{fancyhdr}

\usepackage[T1]{fontenc}
\usepackage{graphicx}
\usepackage{booktabs}
\usepackage{hyperref}
\usepackage{orcidlink}
\usepackage[most]{tcolorbox}

\tcbset{
  colframe=black!40,
  colback=white,
  fonttitle=\bfseries,
  boxrule=0.5pt,
  arc=3pt,
  left=6pt,
  right=6pt,
  top=4pt,
  bottom=4pt
}

\tcbset{
  colback=gray!5,        
  colframe=gray!50,      
  boxrule=0.4pt,         
  arc=2pt,               
  left=6pt, right=6pt, top=4pt, bottom=4pt, 
  fonttitle=\bfseries,   
}

\newtcolorbox{singleRQ}[1]{
  colback=gray!5,
  colframe=gray!50,
  boxrule=0.4pt,
  arc=2pt,
  fonttitle=\bfseries,
  title={#1}
}

\begin{document}

\title{Supporting Stakeholder Requirements Expression with LLM Revisions: An Empirical Evaluation}
\titlerunning{LLMs for Supporting Stakeholder Requirements Expression}
\author{Michael Mircea\orcidlink{0009-0007-6783-4981} \and Emre Gevrek \and Elisa Schmid\orcidlink{0009-0006-2498-9986} \and Kurt Schneider\orcidlink{0000-0002-7456-8323}}
\authorrunning{M. Mircea et al.}

\institute{Leibniz Universität, Welfengarten 1, 30167 Hanover, Germany \\
\email{\{michael.mircea, elisa.schmid, kurt.schneider\}@inf.uni-hannover.de}}
\maketitle              
\begin{abstract}

\textbf{[Context and Motivation]}
Stakeholders often struggle to accurately express their requirements due to articulation barriers arising from limited domain knowledge or from cognitive constraints. This can cause misalignment between expressed and intended requirements, complicating elicitation and validation. \textbf{[Question/Problem]} Traditional elicitation techniques, such as interviews and follow-up sessions, are time-consuming and risk distorting stakeholders’ original intent across iterations. Large Language Models (LLMs) can infer user intentions from context, suggesting potential for assisting stakeholders in expressing their needs. This raises the questions of (i) how effectively LLMs can support requirement expression and (ii) whether such support benefits stakeholders with limited domain expertise.
\textbf{[Principal Ideas/Results]} We conducted a study with 26 participants who produced 130 requirement statements. Each participant first expressed requirements unaided, then evaluated LLM-generated revisions tailored to their context. Participants rated LLM revisions significantly higher than their original statements across all dimensions—\textit{alignment with intent, readability, reasoning}, and \textit{unambiguity}. Qualitative feedback further showed that LLM revisions often surfaced tacit details stakeholders considered important and helped them better understand their own requirements.
\textbf{[Contribution]} We present and evaluate a stakeholder-centered approach that leverages LLMs as articulation aids in requirements elicitation and validation. Our results show that LLM-assisted reformulation improves perceived completeness, clarity, and alignment of requirements. By keeping stakeholders in the validation loop, this approach promotes responsible and trustworthy use of AI in Requirements Engineering.

\end{abstract}
\section{Introduction}
Eliciting and validating stakeholder needs is a central challenge in Requirements Engineering (RE). These tasks are inherently complex because cognitive and social factors hinder stakeholders from accurately expressing their intentions, including memory limitations, societal pressures, and articulation barriers~\cite{tourangeau2000psychology}. While retrieval and reporting barriers arise from limited recall and context dependence, articulation barriers often stem from tacit knowledge or vocabulary mismatches between stakeholders and engineers~\cite{tourangeau2000psychology}. This challenge may be particularly pronounced among stakeholders with limited technical or domain knowledge. Consequently, requirements must be elicited through interactive dialogue rather than simply gathered from stakeholder statements~\cite{zowghi2005requirements}, to enable clarification and shared understanding. Current approaches such as interviews rely on communication-intensive exchanges in which RE experts uncover stakeholders’ underlying needs~\cite{zowghi2005requirements}. Effective elicitation, however, requires substantial domain understanding; without it, resulting artifacts may be low in quality, misaligned with stakeholder intent, or lacking clarity and reasoning~\cite{kustiawan2023user}. To mitigate this, elicitation is often conducted iteratively across multiple sessions~\cite{zowghi2005requirements}, allowing practitioners to contextualize and refine stakeholder input into artifacts such as user stories~\cite{lucassen2016use}. These artifacts form the basis for validation and development but make multi-session formats time-intensive and prone to misalignment, as delays can cause stakeholders to forget or reinterpret their original intentions.

Large Language Models (LLMs) may be able to bridge this gap due to their strong natural language capabilities and domain understanding. Beyond the current evidence of LLM ability to simulate empathetic behavior~\cite{sorin2024large}, existing work explicitly shows that the in-context learning capabilities of LLMs can be effective at identifying user intent~\cite{rodriguez2024intentgpt}. Given sufficient stakeholder context, they may accurately express needs from the stakeholder’s perspective, reducing articulation barriers. Their efficiency also enables immediate validation by the stakeholder, supporting low-latency iteration within a single elicitation session.

User Stories are a popular requirements artifact~\cite{lucassen2016use}, which provide an effective medium for this collaboration, as their structured yet accessible format bridges the gap between stakeholders and developers. Each user story captures stakeholder, functionality and rationale, in a form understandable to both parties. This dual interpretability allows LLM-generated user stories to be validated directly by stakeholders and subsequently used in development without reformulation, minimizing the risk of misinterpretation between elicitation and implementation~\cite{lucassen2016use}.

In this paper, we investigate if the capabilities of LLMs to infer user intent and tacit knowledge can assist stakeholders in expressing software requirements more effectively, more specifically to create pre-validated, high quality user stories aligned with stakeholder intent. We propose a collaborative elicitation approach in which LLMs help stakeholders refine and understand their needs in real time. This raises the following research questions:

\begin{singleRQ}{Research Questions}
\begin{itemize}
    \item \textbf{RQ1:} How effective are LLMs in assisting stakeholders in expressing their requirements?
    \item \textbf{RQ2:} Is this expression support particularly beneficial for stakeholders with limited domain expertise?
\end{itemize}
\end{singleRQ}

\section{Related Work}
There is a rapidly growing body of research exploring the use of Large Language Models in Requirements Engineering (LLM4RE)~\cite{zadenoori2025large,cheng2024generative}. While LLMs are tested for various tasks, the nature of the evaluation is most commonly focused around automation, rather than collaborative systems~\cite{zadenoori2025large}. Given the well-documented risks of hallucination and misalignment~\cite{cheng2024generative}, recent works emphasize the need for Human-in-the-Loop (HITL) approaches in both the design and evaluation of LLM-based systems~\cite{zadenoori2025large,vogelsang2024specifications}. Our work aligns with this direction through an AI-in-the-Loop perspective~\cite{natarajan2025human}, where the human drives the elicitation process and the AI assists through contextual reformulation and clarification. This approach aims to counteract hallucination concerns by maintaining stakeholder agency while reducing cognitive and articulation barriers.

Several studies have examined the potential of LLMs for elicitation and specification tasks. Ronanki et al.~\cite{ronanki2023investigating} compared requirements for trustworthy AI generated by ChatGPT with those written by RE experts. Other expert reviewers rated the LLM-generated requirements as acceptable to high across criteria such as atomicity, consistency, and correctness, but the authors stressed that true requirements must originate from or be validated by the customer, not RE experts. Similarly, Santos et al.~\cite{santosuser} and Akin et al.~\cite{akin2024well} assessed ChatGPT’s ability to generate user stories using frameworks such as QUS and INVEST. Both found that ChatGPT could produce user stories of comparable formal quality to those written by humans, though their evaluations were conducted by RE experts rather than stakeholders. Consequently, these studies primarily assessed adherence to structural quality metrics rather than alignment with stakeholder intent or validity of content. Hymel and Johnson~\cite{hymel2025analysis} extended this line of work by comparing LLM- and human-generated requirements for fifty participant-submitted project ideas. Each idea was transformed into requirements by both a human expert and GPT-4, and the originating participants rated both versions for alignment and completeness. LLM-generated requirements received higher alignment scores and comparable completeness, demonstrating that LLMs can efficiently produce coherent first drafts of requirements.

Collectively, existing studies treat LLMs as independent producers of requirements whose outputs are later validated by others (often not stakeholders). In contrast, our work examines LLMs as stakeholder-centered articulation aids that assist in reformulating and clarifying requirements while keeping the stakeholder as the final validator.
Rather than assessing LLM performance from an expert standpoint, we focus on how LLM-assisted reformulations improve stakeholders’ ability to express, understand, and ultimately validate their own requirements.
This perspective positions LLMs not as autonomous requirement generators, but as supportive systems that help bridge articulation gaps and foster more inclusive, human-centered elicitation practices. To promote replicability, we provide our raw data as well as the full prompt in our supplementary material~\footnote{\url{https://figshare.com/s/ba3cd02a6b69a49846f7}}.

\section{Study Design}
This study examines how large language models (LLMs) can support stakeholders in articulating software requirements by inferring intent and improving clarity and completeness. Unlike approaches focused on automating requirement generation~\cite{akin2024well,santosuser}, we explore LLMs as collaborative aids that refine and reformulate stakeholder statements. To enhance external validity, we selected a specific software family, integrated development environments (IDEs), and involved real users as stakeholders. Since only stakeholders themselves can judge whether a generated requirement accurately represents their needs, they directly validated the LLM revisions. To analyze effects across domain familiarity, both novice and experienced users were included. In total, twenty-six participants were recruited through convenience sampling, comprising software engineering students and professionals from Germany with varying levels of technical expertise.

\subsection{Methodology}
The study was conducted as a guided survey with a moderator who ensured participant understanding while minimizing influence. \autoref{fig:methodology} illustrates the overall process. Participants first provided informed consent and were briefed on the study’s purpose (including the use of AI) and data protection measures. The LLM used was a GDPR-compliant, data-isolated instance of GPT-4o. The survey consisted of three phases:
\begin{figure}[htb]
    \centering
    \includegraphics[width=1\textwidth,
            trim=15 10 20 0, clip]{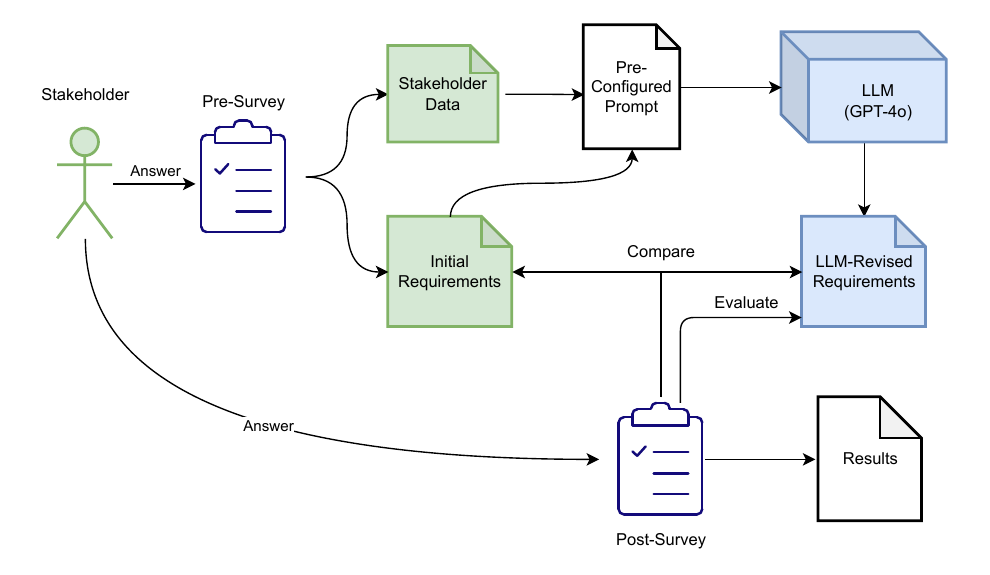}
    \caption{Methodology of our study design. Stakeholder and generated artifacts are colored green, LLM and generated artifacts are colored blue.}
    \label{fig:methodology}
\end{figure}

\begin{enumerate}
    \item \textbf{Pre-survey:} Participants provided contextual data and initial requirements.
    \item \textbf{LLM intervention:} Moderator inserted stakeholder attributes and requirements into a structured prompt template, generating revised requirements.
    \item \textbf{Post-survey:} Participants evaluated the LLM revisions and compared them against their original statements.
\end{enumerate}

\subsubsection{Pre-Survey}
The pre-survey consisted of two parts:
\begin{enumerate}
    \item \textbf{Gathering stakeholder attributes:} Participants reported demographic and contextual information to enable personalized prompt conditioning. Collected variables included age (binned in five-year intervals), gender, education, work experience, and IDE experience (described in natural language).
    \item \textbf{Task instruction and initial requirements:} Participants formulated five requirements for a new IDE using the user story format (“As a [user], I want [goal] so that [benefit]”). This was to allow for fair comparison with LLM-revisions. The moderator ensured consistent format adherence.
\end{enumerate}

\subsubsection{LLM Intervention}
Prompt design followed Google’s Prompt Engineering Whitepaper~\cite{boonstra2025promptengineering}. Iterative pilot testing led to a modular template parameterized with stakeholder data from the pre-survey. This ensured transparency, reusability, and reproducibility. The conceptual components of the final prompt are summarized below (the full prompt is included in our supplementary material).

\tcbset{
  boxrule=0.3pt,                
  arc=3pt,                      
  left=4pt, right=4pt, top=3pt, bottom=3pt, 
  fonttitle=\bfseries\small,    
  fontupper=\small,             
  before skip=5pt, after skip=5pt, 
  enhanced,                     
  coltitle=white,               
}

\begin{tcolorbox}[colback=blue!1,
  colframe=blue!25!black,
  title=Definition of Variables,
  boxed title style={colback=blue!35!white,colframe=blue!25!black}]
Defines variables from the pre-survey for use as contextual placeholders (e.g., \texttt{{age}}, \texttt{{work experience}}), ensuring consistent and efficient prompt contextualization.
\end{tcolorbox}

\begin{tcolorbox}[colback=teal!1,
  colframe=teal!25!black,
  title=Role Prompting,
  boxed title style={colback=teal!35!white,colframe=teal!25!black}]
Specifies the LLM’s assumed role to shape tone and reasoning style. The model acted as a \textit{Product Owner} responsible for articulating and refining user stories.
\end{tcolorbox}

\begin{tcolorbox}[colback=green!1,
  colframe=green!25!black,
  title=Contextual Prompting,
  boxed title style={colback=green!35!white,colframe=green!25!black}]
Provides the LLM with stakeholder background and project context, enabling phrasing from the stakeholder’s perspective with the clarity of a \textit{Product Owner}.
\end{tcolorbox}

\begin{tcolorbox}[colback=orange!1,
  colframe=orange!25!black,
  title=Task Instruction,
  boxed title style={colback=orange!35!white,colframe=orange!25!black}]
Instructs the LLM to revise stakeholder requirements for clarity, completeness, and understandability, adapting detail and vocabulary to the given persona.
\end{tcolorbox}

\begin{tcolorbox}[colback=purple!1,
  colframe=purple!25!black,
  title=Specified Output,
  boxed title style={colback=purple!35!white,colframe=purple!25!black}]
Constrains output format (“As a [user], I want [goal] so that [benefit]”).
\end{tcolorbox}

\begin{tcolorbox}[colback=red!1,
  colframe=red!25!black,
  title=One-Shot Example,
  boxed title style={colback=red!35!white,colframe=red!25!black}]
Includes a sample input–output pair to demonstrate the expected reasoning process and guide the model’s response style.
\end{tcolorbox}

\subsubsection{Post-Survey}
The post-survey captured both quantitative and qualitative feedback. We determined ``effectiveness'' as used in \textbf{RQ1} (``How effective are LLMs in assisting stakeholders in expressing their requirements?'') based on a systematic literature review~\cite{kustiawan2023user} of user stories and their most common defects, as well as further aspects regarding perceived improvements. To measure effectiveness, participants compared each LLM-revised requirement with its original version on a five-point semantic differential scale (much worse → much better) across the following dimension:
\begin{singleRQ}{Set 1: Comparison of original statements and revisions}
\begin{itemize}
    \item \textbf{Alignment}: Representation of stakeholder intent.
    \item \textbf{Readability}: Ease of understanding.
    \item \textbf{Reasoning}: Clarity of rationale behind the requirement.
    \item \textbf{Unambiguity}: Degree to which the requirement avoids ambiguity.
\end{itemize}
\end{singleRQ}

Additionally, participants answered binary questions assessing deeper cognitive and reflective effects:
\begin{singleRQ}{Set 2: Evaluation of each revision}
\begin{itemize}
    \item \textbf{Surfacing tacit knowledge (completeness)}: Did the revision introduce aspects they had not explicitly stated, but considered important?
    \item \textbf{Comprehension}: Did the revision help them understand their own needs?
    \item \textbf{Correctness}: Did the revision introduce factual or logical errors?
\end{itemize}
\end{singleRQ}

Finally, participants provided overall impressions of the revisions:

\begin{singleRQ}{Set 3: Stakeholder perception of all revisions}
\begin{itemize}
    \item \textbf{Perceived strengths}: Aspects the stakeholders liked in particular.
    \item \textbf{Perceived weaknesses}: Unfitting or disruptive aspects.
    \item \textbf{Particular omissions}: Any cases, where a revision omitted or obfuscated previously mentioned, important details of a requirement.
\end{itemize}
\end{singleRQ}

\section{Results}

\subsection{Participant Overview}
A total of 26 participants took part in the study, resulting in 130 paired comparisons. For correlation analysis, the first two authors independently classified participants’ experience levels into two groups based on their self-reported IDE usage and work experience: 16 were categorized as \textit{low-experience} and 10 as \textit{high-experience} users.

\subsection{Quantitative Results: All Participants}

\subsubsection{Comparison of Original and LLM-Revised Requirements}
\begin{figure}[htb]
    \centering
    \includegraphics[width=1\linewidth]{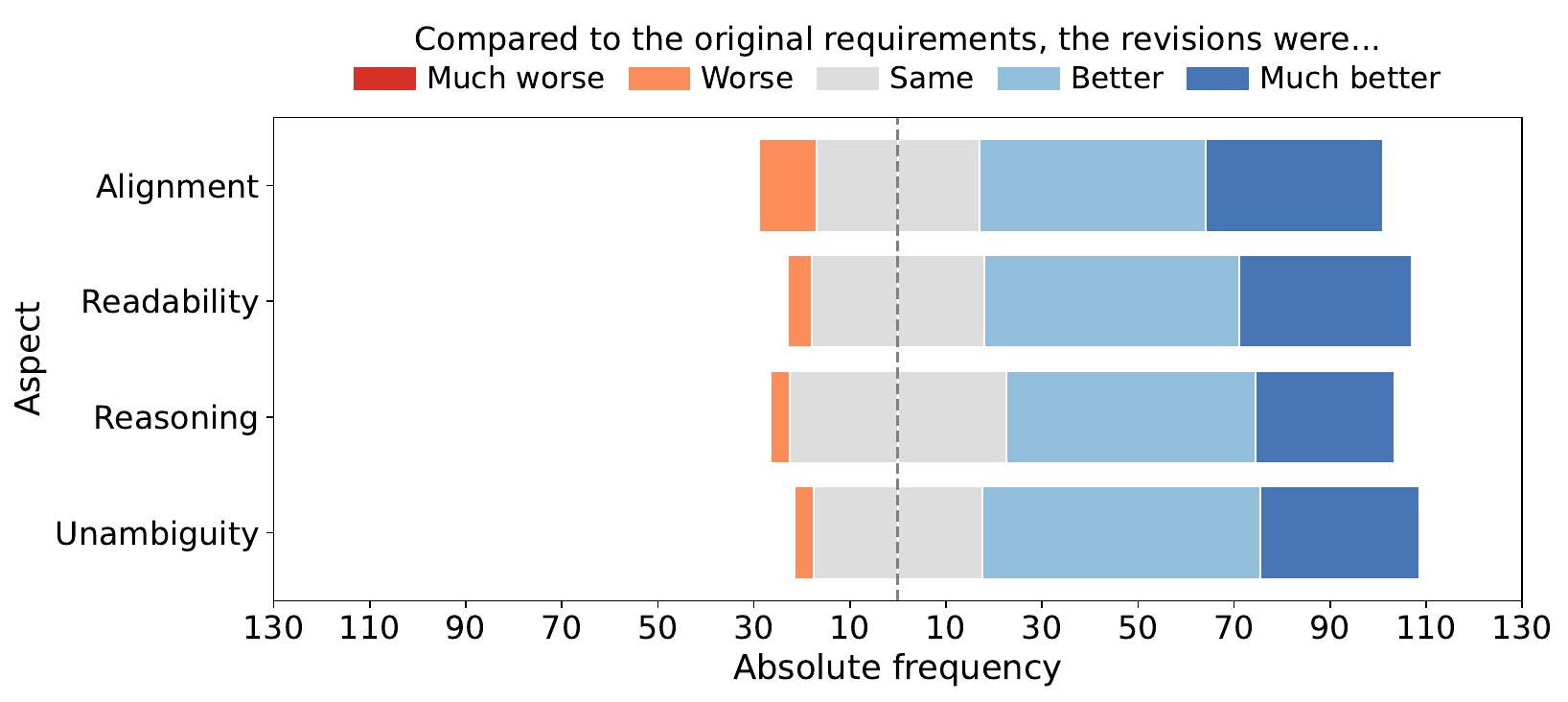}
    \caption{Participant ratings comparing LLM-revised to original requirements in four dimensions. Bars show the proportion of ratings on a five-point scale from LLM-revisions being “much worse” to “much better.”}
    \label{fig:bar_chart_all}
\end{figure}

\autoref{fig:bar_chart_all} illustrates the aggregated participant ratings comparing the LLM-revised requirements to their original statements across four dimensions: \textit{Alignment}, \textit{Readability}, \textit{Reasoning}, and \textit{Unambiguity}. 

Across all dimensions, LLM-revised requirements were rated higher than their original counterparts. 
While \autoref{fig:bar_chart_all} presents all individual ratings ($N=130$ requirement pairs), the statistical analysis was performed on aggregated data per participant ($N=26$), with the results presented in \autoref{tab:wilcoxon_results}. For each dimension, every participant’s five ratings were summarized using their median score, resulting in one representative value per dimension and participant. This approach mitigates dependence between repeated measures and provides a participant-level view of the results. In the aggregated data, no participant rated any dimension below the neutral midpoint of the scale (“equal”), and most medians corresponded to “better” ($4$). 
Wilcoxon signed-rank tests against the neutral value confirmed that these improvements were statistically significant across all dimensions ($p < 10^{-5}$), with large effect sizes ($r = .876$–$.878$). 
To account for multiple testing across the four evaluated dimensions, a Bonferroni correction was applied ($\alpha_\text{adj} = 0.0125$). 
All effects remained well below this threshold, indicating robust and consistent improvements across all quality aspects.

\begin{table}[htb]
\centering
\caption{Wilcoxon signed-rank test results comparing original and LLM-revised requirements (aggregated per participant, $N = 26$). Reported are median ratings, test statistic ($W$), $p$-values, and effect sizes ($r$).}
\label{tab:wilcoxon_results}
\begin{tabular*}{\textwidth}{l@{\extracolsep{\fill}}cccc}
\toprule
\textbf{Dimension} & \textbf{Median} & \textbf{$W$} & \textbf{$p$} & \textbf{$r$} \\
\midrule
Alignment & Better & 0.0 & $< .001$ & .878 \\
Readability & Better & 0.0 & $< .001$ & .878 \\
Reasoning & Better & 0.0 & $< .001$ & .878 \\
Unambiguity & Better & 0.0 & $< .001$ & .876 \\
\bottomrule
\end{tabular*}
\end{table}

It is notable how similar the results of the statistical analysis are across dimensions. This is, because aggregating by median naturally smooths out within-participant variation. Therefore, we additionally performed an exploratory analysis treating all 130 samples as independent observations.
While this approach underestimates $p$-values due to non-independence ($p < 10^{-15}$), it revealed a subtle pattern across dimensions: \textit{Readability}, \textit{Reasoning}, and \textit{Unambiguity} showed similarly strong effects ($r \approx .85$), whereas \textit{Alignment} was slightly lower ($r = .76$).
This trend reflects the distribution in \autoref{fig:bar_chart_all} more accurately, where participants rated alignment “worse” more often compared to other dimensions.

\subsubsection{Supplementary Yes/No Responses}
\begin{figure}[htb]
    \centering
    \includegraphics[width=\textwidth]{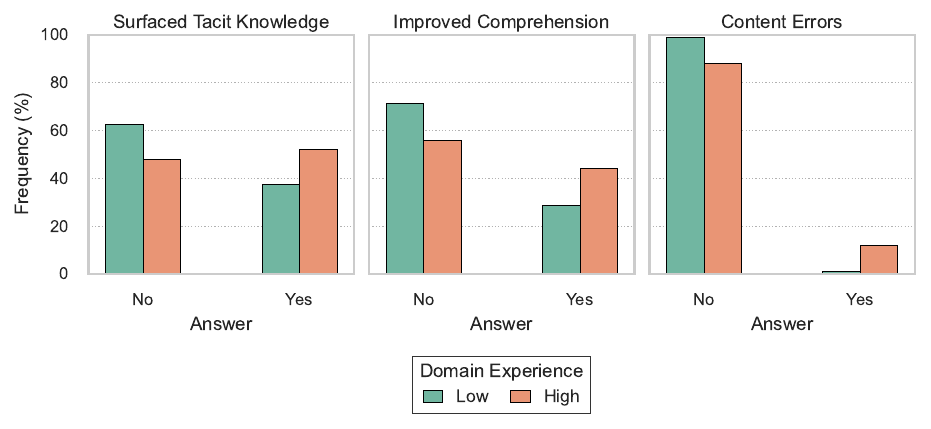}
    \caption[
        Comparison of yes/no responses by domain experience.
    ]{
        Proportion of perceived improvements or issues in LLM-revised requirements ($N = 130$).  Each bar shows the percentage of participants in 
        the low- and high-experience groups answering “Yes” or “No.” 
    }
    \label{fig:yes_no_responses}
\end{figure}

Figure~\ref{fig:yes_no_responses} summarizes participants’ binary responses on whether the LLM revisions:
(1) surfaced tacit details they had forgotten, 
(2) improved their understanding of the requirement, or 
(3) introduced any factual or logical errors. 
The results are visually divided between low- and high-experience participant responses.

In 43\% of all evaluations, participants indicated that the LLM revisions correctly surfaced additional aspects they had not originally mentioned, while 35\% stated that the revisions improved their understanding of the underlying requirement. Only 5\% of revisions (7 out of 130) were perceived to introduce factual or logical inaccuracies. A closer look at the responses suggests that high-experience participants more frequently recognized added details and improved comprehension, but also identified a greater number of content-related errors (six cases compared to one among low-experience participants).

\subsection{Qualitative Analysis}
To complement the quantitative data, responses to the concluding questions were analyzed thematically through open coding of the free-text answers.

\subsubsection{Perceived Strengths}
Participants commented on strengths of the revised requirements. Table~\ref{tab:qual_strengths} summarizes the resulting themes and their frequency of mention.
Most comments emphasized improvements in clarity, readability, and linguistic quality, with several participants also noting that the revisions added useful details or supported their understanding of the requirements.

\begin{table}[htb]
\centering
\caption[
    Perceived strengths of LLM-revised requirements.
]{
    Perceived strengths of LLM-revised requirements.
}
\label{tab:qual_strengths}
\begin{tabular*}{\textwidth}{@{\extracolsep{\fill}}ll}
\toprule
\textbf{Theme} & \textbf{Mentions} \\
\midrule
Clearer and more precise formulations & 13 \\
Improved readability, sentence structure, and word choice & 13 \\
Additional details and elaborations & 7 \\
Support in understanding and expressing requirements & 5 \\
Reduced misunderstandings and clarified meaning & 4 \\
More professional and polished writing style & 4 \\
Preserved the original idea and intent & 2 \\
\bottomrule
\end{tabular*}
\end{table}

\subsubsection{Perceived Weaknesses}
Participants were also invited to comment on any aspects of the LLM-generated revisions they found distracting or unhelpful. Thematic analysis of these free-text responses revealed that most participants explicitly stated that they did not perceive the revisions as problematic, while a minority reported issues related to loss of meaning, complexity, or over-elaboration.
Table~\ref{tab:qual_weaknesses} summarizes the main themes and their frequency of mention.

\begin{table}[htb]
\centering
\caption[
    Perceived weaknesses of LLM-revised requirements.
]{
    Perceived weaknesses of LLM-revised requirements.
}
\label{tab:qual_weaknesses}
\begin{tabular*}{\textwidth}{@{\extracolsep{\fill}}ll}
\toprule
\textbf{Theme} & \textbf{Mentions} \\
\midrule
No issues perceived & 17 \\
Loss of meaning, focus, or thematic deviation & 9 \\
Overly elaborate or unnecessarily complex phrasing & 3 \\
Ambiguous or unclear wording & 2 \\
Minimal change or redundant adjustment & 1 \\
\bottomrule
\end{tabular*}
\end{table}

\subsubsection{Lost or Omitted Details}
Finally, participants were asked whether any important details or meanings were lost in the LLM-generated revisions. Most participants (20 out of 26) reported no loss of information in any of their five evaluated revisions.
Among the few who did, two distinct types of meaning loss were described: one participant noted that certain words meant to be preserved were replaced or rephrased, resulting in a subtle linguistic loss of meaning, while another indicated that the revision missed the core content of their requirement.

\subsection{Comparison by Experience Level}
\begin{figure}[htb]
    \centering
    \includegraphics[width=\textwidth]{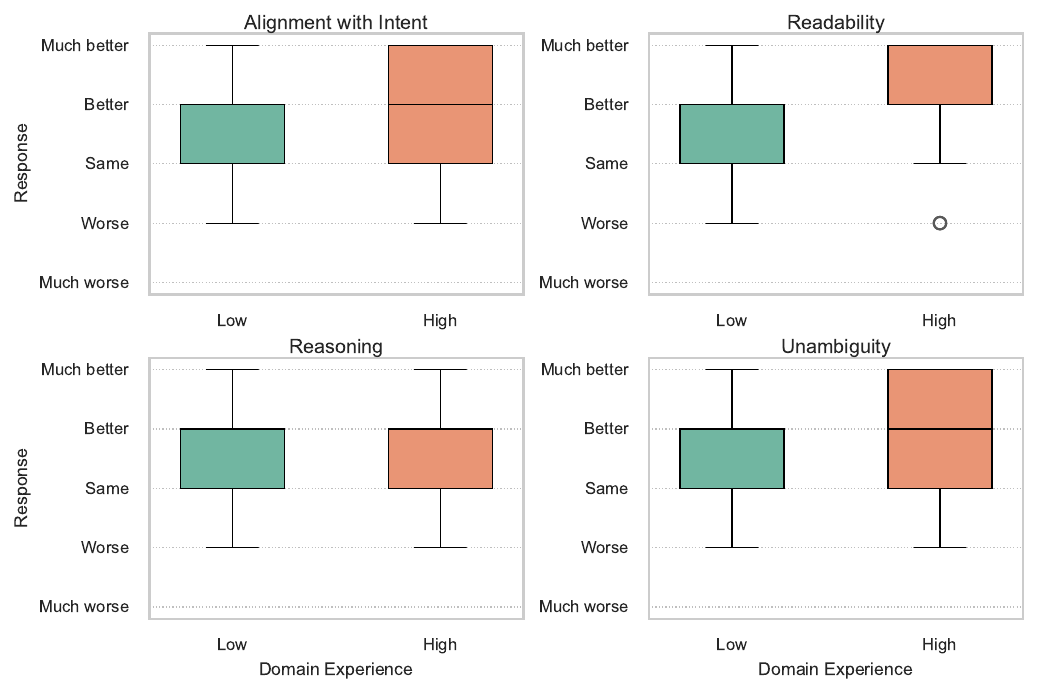}
    \caption[
        Comparison of low- and high-experience participants across four evaluation dimensions.
    ]{
        Comparison of responses from low- and high-experience stakeholders.
    }
    \label{fig:boxplots_experience_comparison}
\end{figure}

Figure~\ref{fig:boxplots_experience_comparison} visualizes the distribution of ratings across low- and high-experience participants for each evaluated dimension, while the corresponding statistical results are summarized in Table~\ref{tab:group_comparison_results}.
Overall, the boxplots show largely similar response patterns between experience groups, with median ratings consistently favoring the LLM-revised requirements across all dimensions.
A Mann--Whitney~U test revealed a small raw difference for readability (raw $p_{raw} = .045$), suggesting that more experienced stakeholders may have perceived stronger linguistic improvements.
However, this difference did not remain significant after applying Holm’s correction for multiple comparisons ($p_\text{adj} = .18$).
All other dimensions showed no statistically significant differences between groups.
Complementary Chi-square tests on the full scale distributions confirmed this pattern, indicating no meaningful distributional shifts.

\begin{table}[htb]
\centering
\caption[
    Group comparison between low- and high-experience participants (Holm correction).
]{
    Mann-Whitney~U and Chi-square test results comparing 
    experience groups across the four evaluation dimensions.  
    Holm-adjusted $p$-values ($p_\text{adj}$) for $k=4$. 
}
\label{tab:group_comparison_results}
\begin{tabular*}{\textwidth}{@{\extracolsep{\fill}}lcccccc}
\toprule
\textbf{Dimension} & \textbf{$U$} & \textbf{$p_\text{MWU}$} & \textbf{$p_\text{MWU, adj}$} 
& \textbf{$\chi^2$} & \textbf{$p_\text{Chi}$} & \textbf{$p_\text{Chi, adj}$} \\
\midrule
Alignment with Intent & 1903.5 & .630 & .630 & 4.10 & .251 & .502 \\
Readability & 1604.5 & \textbf{.045} & .180 & 5.01 & .171 & .684 \\
Reasoning & 1984.5 & .939 & .939 & 0.42 & .936 & .936 \\
Unambiguity & 1813.0 & .340 & .680 & 3.98 & .264 & .528 \\
\bottomrule
\end{tabular*}
\end{table}

\section{Discussion}
\subsection{Interpretation of Findings}
The results demonstrate that LLM-assisted revisions substantially improved the overall quality of stakeholder requirements. Participants rated the revised statements significantly higher across all four evaluated dimensions, with large and consistent effect sizes at the participant level. Notably, none of the revisions were rated as ‘much worse’ by any participant. While we anticipated linguistic improvements, it was surprising that \textit{Alignment} also increased significantly and that many revisions were reported to improve completeness and enhance understanding. When viewed at the requirement level, the relative pattern across dimensions provides further insight: the strongest improvements appeared in \textit{Readability} ($r \approx .85$), \textit{Reasoning} ($r \approx .85$), and \textit{Unambiguity} ($r \approx .85$), suggesting that current LLMs excel primarily in linguistic refinement, structural clarity, and logical articulation. Nevertheless, \textit{Alignment} ($r = .76$) also showed a strong improvement, indicating that LLM assistance can help stakeholders express their underlying needs more faithfully.

The qualitative results corroborate these findings: participants emphasized clearer phrasing, smoother sentence structure, and helpful elaborations as major strengths, while only a few noted over-elaboration or semantic drift. Most did not perceive any information loss, and when meaning deviations occurred, they were either lexical (word choice) or conceptual (focus shift). These isolated instances of altered meaning highlight the importance of keeping LLM4RE systems within a collaborative validation loop to ensure that humans remain the ultimate authority on requirements correctness and intent.

Differences between low- and high-experience participants were small and statistically non-significant after adjustment for multiple testing, suggesting that the benefits of LLM support are similar across different levels of stakeholder expertise. Interestingly, experienced participants tended to rate the revisions slightly higher and more often reported improved understanding and additional beneficial details. This trend, while contrary to our initial expectation that novices would benefit most from expression support, may suggest that experienced users are better able to recognize nuanced improvements and critically evaluate precision of requirements. This interpretation is reinforced by the fact that the same group also identified many more inaccuracies or errors, indicating heightened sensitivity rather than diminished benefit.

\subsection{Interpretive Limitations}
Although the observed effect sizes were very strong, several interpretive limitations must be considered before deriving implications for RE practice. Further, methodological limitations will be discussed in~\autoref{sec:threats-to-validity}. 

\subsubsection{Domain Specificity and Stakeholder Type}
The study focused on requirements related to IDEs. While this choice ensured that the authors could verify the plausibility of participant statements and LLM revisions, it also represents a highly technical and structured context. All participants were active IDE users, implying familiarity with abstract reasoning and formalized expression. Consequently, the linguistic and cognitive demands in this domain may differ substantially from those in less technical contexts. In stakeholder groups with limited articulation ability (e.g., patients, educators, or elderly users), the effects of LLM support might vary. Further, IDEs are an established software category. The performance of LLMs to infer stakeholder needs may be lower when applying them to novel or innovative projects. Future studies should therefore examine whether similar improvements occur in non-technical or multidisciplinary RE settings.

\subsubsection{Evaluation Design and Perceived Quality}
Our evaluation compared pairs of user stories. While this ensured fair and controlled comparisons, it may not fully reflect real-world elicitation dynamics, which are often more unstructured and conversational. Moreover, the quality judgments were based on stakeholder perception rather than external expert evaluation. While this focus was intentional and already justified in the earlier sections, we recognize our results capture only perceived improvements in requirement quality. Therefore, our approach does not replace technical analyses. Subsequent RE processes must still ensure factual accuracy, feasibility, and consistency with system goals.

\subsection{Implications for RE Practice}
The findings suggest several implications for how LLMs can be responsibly integrated into RE practice. First, LLMs can serve as effective articulation support during early elicitation activities. Participants consistently rated LLM revisions as clearer, more reasoned, and less ambiguous than their original statements. Combined with the expert-based evaluation from related works~\cite{akin2024well,santosuser}, this indicates that LLMs can help stakeholders formulate high-quality requirements in real time. This capability could be leveraged in interviews, workshops, or novel digital elicitation tools to help stakeholders express their needs more precisely before analyst inspection. Second, the observed improvements in comprehension and completeness highlight LLMs’ potential to act as reflective partners in elicitation. Stakeholders frequently reported that model-generated revisions surfaced tacit details they had not articulated and helped them better understand their own requirements. Embedding such LLM assistance in interactive elicitation processes may therefore promote early validation and richer stakeholder reflection, leading to more accurate and aligned requirements from the start.

However, the occasional semantic drift observed underscores that such support must remain collaborative. LLMs should not replace stakeholders or practitioners in RE processes but rather complement them as a liaison. Keeping both parties in the loop ensures that improved formal quality does not come at the cost of alignment with stakeholder needs or traceability, promoting responsible, human-centered application of AI in Requirements Engineering.

Finally, the findings suggest practical design directions for future RE tools. Since perceived benefits were similar across experience levels, LLM-based assistance could be broadly applicable when tailoring phrasing and feedback depth to stakeholder attributes and expertise. Further, integrating such tools into distributed or asynchronous elicitation formats (i.e. online questionnaires, chat-based interviews or app reviews) could substantially improve requirement clarity and mutual understanding without increasing process overhead.

\section{Threats to Validity}
\label{sec:threats-to-validity}
This section discusses potential threats to validity following Wohlin et al.~\cite{wohlin2012experimentation}.

\subsubsection{Construct Validity}
Our operationalization of “effective” expression support was based on the stakeholders’ own perception of quality and alignment. This intentionally emphasizes subjective assessment over objective correctness, as only stakeholders can judge whether a requirement accurately represents their intent or enhances their understanding. Perceived improvement does not necessarily imply factual correctness or feasibility, however these qualities of LLM-generated user stories were already investigated by related works~\cite{hymel2025analysis,santosuser,akin2024well}.

\subsubsection{Internal Validity}
To guarantee informed consent, participants were informed that the study involved an AI-based technology, which may have influenced their evaluations. Some participants might have rated LLM outputs more favorably due to an \textit{appeal-to-authority} or \textit{social desirability} bias, while others may have rated them more critically due to skepticism toward AI. Since participants evaluated revisions of their own requirements, self-evaluation factors could also have influenced effects. Furthermore, the LLM’s role of \textit{Product Owner} may have shaped stylistic outcomes associated with professional quality. We counteracted this by keeping the same user story structure for both original and revised statements. Finally, our prompts included different context information, including stakeholder attributes, to tailor the model’s phrasing to an adequate level of  complexity. While this aligns with guidelines~\cite{boonstra2025promptengineering}, we did not empirically verify the effect of these attributes against a zero-shot baseline. Therefore, the impact of individual prompt components remains an unverified assumption that may have influenced performance.

\subsubsection{Conclusion Validity}
The modest participant sample ($N = 26$, 16 low, 10 high) limits statistical power for subgroup analyses. To address dependence among repeated measures, we aggregated ratings per participant before hypothesis testing, which reduced within-subject variation. However, while this leads to more accurate statistical measures, aggregating by median smooths out some nuances in the data. We therefore additionally performed an exploratory reanalysis using all individual ratings which produced comparable patterns with more nuanced insights on effect sizes, suggesting robustness.

\subsubsection{External Validity}
The study was conducted in a realistic application context but limited to users of IDEs. This represents a technically literate population that may be more articulate and comfortable with structured reasoning than typical end users. Consequently, the observed effects may not generalize to less technical domains such as healthcare or education, where articulation barriers and domain-specific language differ substantially. Moreover, the study was conducted with a single, GDPR-compliant instance of GPT-4o. Different model versions or prompts could yield different results.

\section{Conclusion and Future Work}
This work investigated how LLMs can support stakeholders in expressing their software requirements more effectively. Using a stakeholder-centered design, we found that LLM-assisted revisions were consistently rated higher than original stakeholder statements across all evaluated dimensions (\textit{Alignment, Readability, Reasoning}, and \textit{Unambiguity}). These results indicate that LLMs can help bridge articulation barriers, leading to clearer and more complete requirements, while preserving stakeholder intent. Qualitative findings further suggest that such assistance can help stakeholders better understand and thus validate their own requirements. Our results highlight the potential of integrating LLMs as articulation and validation aids in early requirements elicitation, particularly in interviews, workshops, or digital elicitation tools. However, occasional meaning drift emphasizes that LLMs should complement, not replace, humans. Keeping stakeholders at the center of the validation loop remains essential for trustworthy and responsible use of LLM4RE systems and practices.

Future work should examine the generalizability of these results in less technical or multidisciplinary domains, explore the impact of contextual attributes in prompting, and evaluate integration of LLM expression support in interactive RE practices or tools. Further research is also needed to determine how well such systems can adapt their phrasing complexity, detail level, and clarification strategies to different stakeholder profiles and contexts.

\begin{credits}

\subsubsection{\discintname}
The authors have no competing interests to declare that are relevant to the content of this article.

\subsubsection{Data Availability Statement}
Our supplementary material is available at \url{https://figshare.com/s/ba3cd02a6b69a49846f7}.
\end{credits}

\bibliographystyle{splncs04}  
\bibliography{references}     
\end{document}